\documentclass[twocolumn,aps,prd]{revtex4}
 \usepackage{graphicx,amssymb}
 \usepackage[utf8]{inputenc} 
 
\usepackage[urlcolor=blue,colorlinks,breaklinks=true]{hyperref}

\PassOptionsToPackage{obeyspaces,spaces,hyphens}{url}
 
 \usepackage{amsmath}

\begin{document}
 	\def\half{{1\over2}}
 	\def\shalf{\textstyle{{1\over2}}}
 	
 	\newcommand\lsim{\mathrel{\rlap{\lower4pt\hbox{\hskip1pt$\sim$}}
 			\raise1pt\hbox{$<$}}}
 	\newcommand\gsim{\mathrel{\rlap{\lower4pt\hbox{\hskip1pt$\sim$}}
 			\raise1pt\hbox{$>$}}}

\newcommand{\be}{\begin{equation}}
\newcommand{\ee}{\end{equation}}
\newcommand{\bq}{\begin{eqnarray}}
\newcommand{\eq}{\end{eqnarray}}
 	
\title{Parameter-free velocity-dependent one-scale model for domain walls}
 	 	
\author{P.P. Avelino}
\email[Electronic address: ]{pedro.avelino@astro.up.pt}
\affiliation{Instituto de Astrof\'{\i}sica e Ci\^encias do Espa{\c c}o, Universidade do Porto, CAUP, Rua das Estrelas, PT4150-762 Porto, Portugal}
\affiliation{Departamento de F\'{\i}sica e Astronomia, Faculdade de Ci\^encias, Universidade do Porto, Rua do Campo Alegre 687, PT4169-007 Porto, Portugal}
\affiliation{School of Physics and Astronomy, University of Birmingham,Birmingham, B15 2TT, United Kingdom}

\date{\today}
\begin{abstract}
We develop a parameter-free velocity-dependent one-scale model for the evolution of the characteristic length $L$ and root-mean-square velocity $\sigma_v$ of standard domain wall networks in homogeneous and isotropic cosmologies. We compare the frictionless scaling solutions predicted by our model, in the context of cosmological models having a power law evolution of the scale factor $a$ as a function of the cosmic time $t$ ($a \propto t^\lambda$, $0< \lambda < 1$), with the corresponding results obtained using field theory numerical simulations. We show that they agree well (within a few $\%$) for root-mean-square velocities $\sigma_v$ smaller than $0.2 \, c$ ($\lambda \ge 0.9$), where $c$ is the speed of light in vacuum, but significant discrepancies occur for larger values of $\sigma_v$ (smaller values of $\lambda$). We identify problems with the determination of $L$ and $\sigma_v$ from numerical field theory simulations which might potentially be responsible for these discrepancies.
\end{abstract}

\maketitle
 	
\section{Introduction}
\label{sec:intr}

The formation of cosmic defect networks in the early universe as a result of symmetry breaking phase transitions is a generic prediction of cosmo-particle physics \cite{Kibble:1976sj}. These networks may in general survive until late cosmological times, leaving behind distinct observational signatures \cite{2000csot.book.....V,Ade:2013xla}.  Most of the recent work on cosmic defects has focused on cosmic strings, in particular due to the realization that they may provide a window into string theory \cite{Copeland:2003bj, Dvali:2003zj, Sarangi:2002yt,Sousa:2016ggw}. However, other types of defects may also play a crucial role in cosmology. Among these, domain walls have attracted less attention, essentially due to the tight observational bounds which require domain walls to be either extremely light or to have decayed long before the present epoch (see \cite{Larsson:1996sp,Hindmarsh:1996xv,Avelino:2008qy,Avelino:2008mh,Hiramatsu:2013qaa,Correia:2014kqa,Kitajima:2015nla,Krajewski:2016vbr,Saikawa:2017hiv} for a discussion of the evolution and cosmological consequences of unstable domain walls). Although, contrary to earlier expectations \cite{Bucher:1998mh,Battye:1999eq}, the possibility of a major domain wall contribution to the current acceleration of the universe has been ruled out both on theoretical and observational grounds \cite{PinaAvelino:2006ia}, there is still room for significant astrophysical and cosmological signatures of light domain walls \cite{Sousa:2015cqa,Lazanu:2015fua}.

In order to obtain accurate predictions of the observational consequences of cosmic defects it is necessary to accurately determine their cosmological evolution, which is usually done resorting to numerical simulations or semi-analytical models. The Velocity-dependent One-Scale (VOS) model provides a statistical description of the large-scale cosmological evolution of defect networks in terms of two macroscopic dynamical variables, the characteristic length of the network $L$ and its root-mean-squared velocity $\sigma_v$. This model, originally proposed with the aim of describing the cosmological evolution of cosmic strings \cite{Martins:1996jp,Martins:2000cs}, has been calibrated using numerical simulations and later extended to account for the dynamics of domain walls \cite{Avelino:2005kn}. More recently, this framework was further extended to describe the macroscopic evolution of relativistic and non-relativistic featureless $p$-branes in $N+1$-dimensional homogeneous and isotropic spacetimes (with $N > p$) in a unified manner \cite{Avelino:2010qf,Avelino:2011ev,Sousa:2011ew,Sousa:2011iu,Avelino:2015kdn}. This unified paradigm has shown to be extremely useful not only in cosmology but also in a variety of other contexts, including condensed matter \cite{Avelino:2010qf} and biology \cite{2012PhRvE..86c1119A}.

The original VOS model has two phenomenological parameters, usually referred to as energy loss and momentum parameters. In \cite{Martins:2016ois} the authors verified that the VOS model is unable to accurately reproduce the cosmological evolution of $L$ and $\sigma_v$ in generic cosmological regimes if these two parameters are assumed to be constant. This motivated the generalization of the VOS model to include four new (constant) additional parameters. The generalized VOS model has been claimed to significantly improve the fit to numerical simulations of domain wall network evolution \cite{Martins:2016ois} at the cost of triplicating the number of phenomenological parameters of the original model. 

In the present paper we shall follow an orthogonal approach and develop a new VOS model for the cosmological dynamics of standard domain walls free from adjustable dynamical parameters. The outline of this paper is as follows. In Sec. \ref{sec2} the evolution of thin featureless spherical and cylindrical domain walls in homogeneous and isotropic universes is described. In Sec. \ref{sec3} we introduce our non-parametric VOS model and compute, as a function of $\lambda$, the associated frictionless scaling solutions for $L$ and $\sigma_v$. A comparison of the predictions of our new VOS model with the results of numerical simulations is made in Sec. \ref{sec4}, including a discussion of the discrepancies and their origin. Finally we conclude in Sec. \ref{sec5}.

\begin{figure}[!htb]
	\centering
	\includegraphics[width=8.0cm]{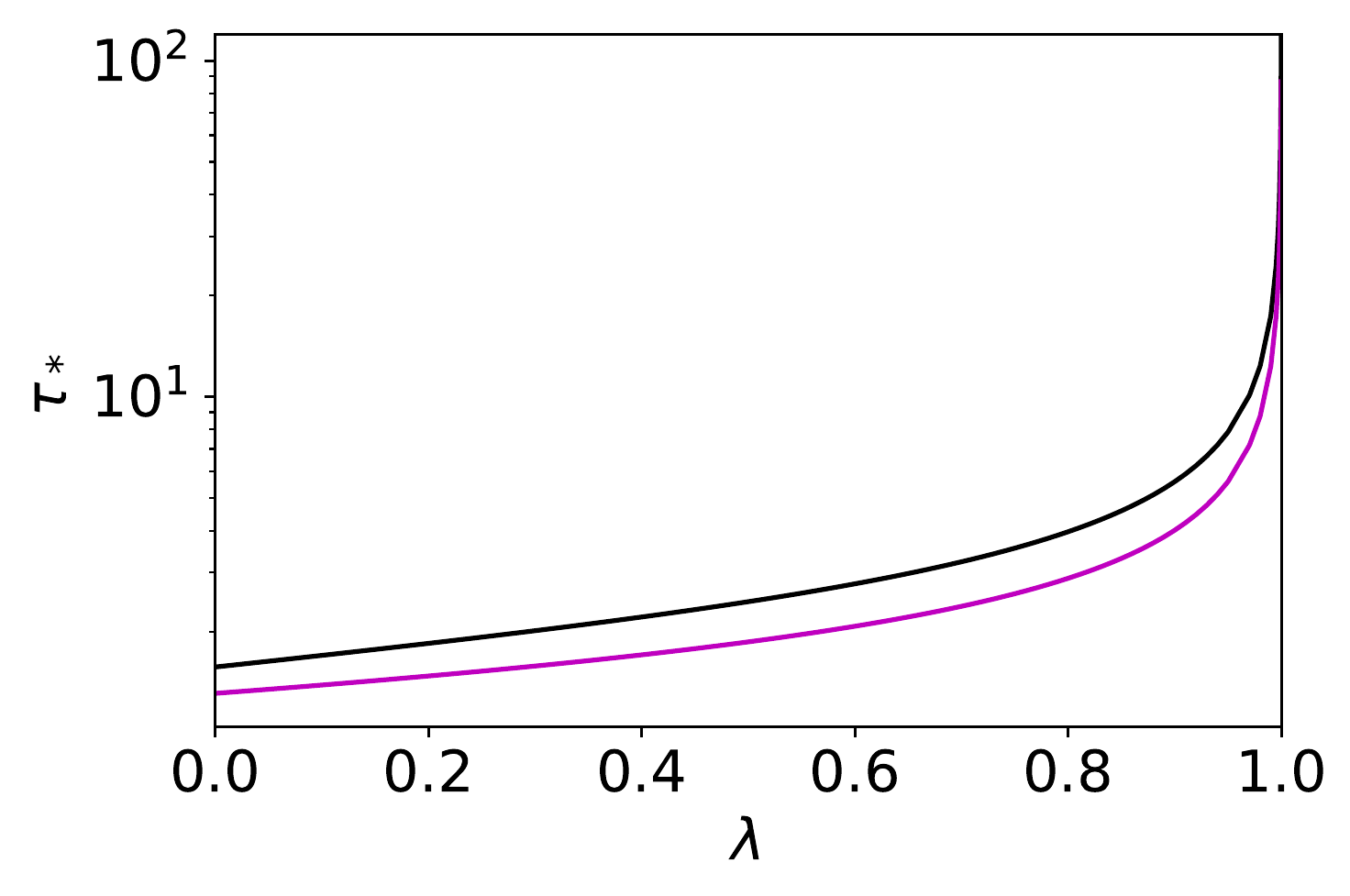}
	\caption{The value of $\tau_* \equiv \eta/q_{i*}$ as a function of  $\lambda$ obtained considering the collapse of cylindrical (upper black line) and spherical (lower magenta line) domain walls. Notice the extremely fast increase of $\tau_*$ with $\lambda$ in the $\lambda \to 1$ limit.}
	\label{fig1}
\end{figure}

We shall use fundamental units with $c=1$, where $c$ is the value of the speed of light in vacuum.

\section{Evolution of spherical and cylindrical  domain walls}
\label{sec2}

Consider a flat $3+1$-dimensional homogeneous and isotropic Friedmann-Lemaitre-Robertson-Walker (FLRW) universe with line element
\be
d s^2 =a^2[\eta] \left(d \eta^2 - d {\bf x} \cdot d {\bf x}  \right) \,,
\ee
where $a$ is the scale factor, $\eta=\int dt/a$ is the conformal time, $t$ is the physical time and $\bf x$ are comoving spatial coordinates. In this paper we shall consider cosmological models that have a power law evolution of the scale factor with the physical time: $a \propto t^\lambda$ with $0<\lambda <1$, so that $a \propto \eta^{\lambda/(1-\lambda)}$.

The world history of an infinitely thin featureless domain wall in a flat expanding FRLW universe can be represented by a three-dimensional world-sheet obeying the usual Goto-Nambu action. In the case of a spherical or cylindrical domain wall the corresponding equations of motion are given by
\be
{\dot q}=-v \,, \qquad {\dot v}+\gamma^{-2}\left(3 {\mathcal H} v - s/q\right)=0\,, 
\label{dwev}
\ee
where a dot denotes a derivative with respect to the conformal time $\eta$, ${\mathcal H}={\dot a}/a > 0$, $q$ is the comoving radius of the wall, $v$ represents its velocity, $\gamma \equiv (1-v^2)^{-1/2}$, and $s=1$ or $s=2$ depending, respectively, on whether the domain wall is cylindrical or spherical. 

\begin{figure}[!htb]
	\centering
	\includegraphics[width=8.0cm]{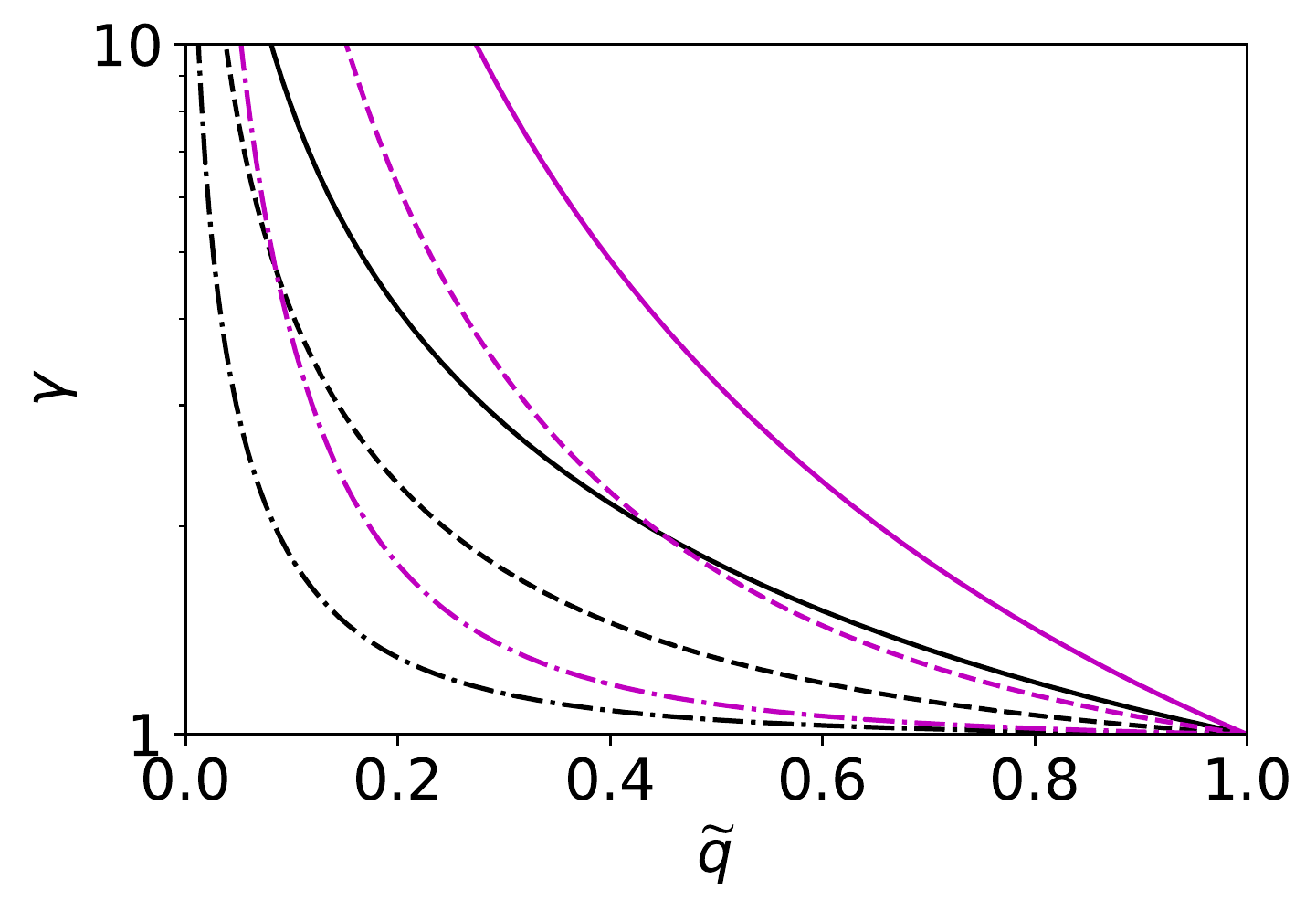}
	\caption{The value of $\gamma$ as a function of $\widetilde q \equiv q/q_i$ obtained considering the evolution of cylindrical (black) and spherical (magenta) domain walls in the context cosmological models with $\lambda=0.1$, $0.5$ and $0.9$ (solid, dashed and dot-dashed lines, respectively). Notice that, for a fixed value of $\widetilde q$, the values of $\gamma$ increase substantially as $\lambda$ approaches zero and are significantly larger for spherical than for cylindrical domain walls.}
	\label{fig2}
\end{figure}

Here, we shall only consider the evolution of domain walls in the comoving radius interval $]0,q_i]$ ($q_i$ represents their initial comoving radius), so that $q$ is always greater or equal to zero. Equation (\ref{dwev}) implies that the intrinsic radius $R \equiv a \gamma^{1/s} q$  of the domain wall evolves 
as \cite{Avelino:2008mv}
\be
{\dot R}= {\mathcal H} R \left(1 - \frac{3}{s} v^2\right)\,.
\label{dwev1}
\ee
Hence, the expansion of the universe is responsible for an increase or decrease of the energy (spherical case) or the energy per unit length (cylindrical case) of the domain wall ($E = 2 \pi s \sigma_{w0} R^s$) depending on whether  $v^2$ is smaller or greater than $s/3$, respectively ($\sigma_{w0}$ is the proper domain wall energy per unit area).

Here, we are particularly interested in the evolution of cylindrical and spherical domain walls starting from rest at the conformal time $\eta_i$ with an initial comoving domain wall radius much larger than the comoving horizon at that initial time ($q_i \gg \eta_i$) --- the initial value of the scale factor shall be normalized to unity ($a_i=1$). Assuming that the domain walls decay upon reaching $q = 0$, at an arbitrary time $\eta \gg \eta_i$ only domain walls with an initial comoving radius larger than a threshold $q_{i*}[\eta]$ survive.

Given a power law expansion with ${\mathcal H} \eta=\lambda/(1-\lambda)$, the equations of motion of cylindrical and spherical domain walls shown in Eq. (\ref{dwev}) are invariant under the transformation $q \to \alpha q$, $\eta \to \alpha \eta$, where $\alpha > 0$ is a constant, leaving the velocity $v$ unchanged. This implies that $v[\eta,q_i]=v[\eta/q_i,1]\equiv  v [\tau]$ and $\gamma[\eta,q_i]=\gamma[\eta/q_i,1]\equiv \gamma [\tau]$, where $\tau=\eta/q_i$. On the other hand $q[\eta,q_i]/q_i= q[\eta/q_i,1] \equiv {\widetilde q}[\tau]$, with ${\widetilde q}[0]=1$ and ${\widetilde q}[\tau_*]=0$. The dependence of the value of $\tau_* \equiv \eta/q_{i*}$ on $\lambda$ is shown in Fig. \ref{fig1} considering the collapse of cylindrical (upper black line) and spherical (lower magenta line) domain walls. Notice that the value of $\tau_*$ increases extremely fast with $\lambda$ for values of $\lambda$ close to unity ($\tau_* \to \infty$ for $\lambda \to 1^-$). Figure \ref{fig2} shows the value of $\gamma$ as a function of ${\widetilde q}$ for cylindrical (black) and spherical (magenta) domain walls, considering cosmological models with $\lambda=0.1$, $0.5$ and $0.9$ (solid, dashed and dot-dashed lines, respectively). Notice that the values of $\gamma$ obtained for a fixed $\widetilde q$ increase substantially as $\lambda$ approaches zero and are significantly larger for spherical than for cylindrical domain walls.

\section{Parameter-free VOS model}
\label{sec3}

Standard domain wall networks are composed of closed spatial domains bounded by domain walls without junctions. The corresponding spatial pattern may be simply described as a hierarchy of smaller domains inside arbitrarily large domains, with the average macroscopic domain wall energy density (over cosmological scales) being inversely proportional to the characteristic length scale $L$ of the network. On the other hand, field theory simulations of domain wall networks evolution have shown that intersections between domain walls occur much less frequently than in the case of cosmic strings. Essentially, domain wall evolution may be described as the successive collapse of increasingly larger domains (the conformal timescale of  domain wall collapse, for a fixed $\lambda$, being roughly proportional to the initial comoving size of the domain walls). Also, thin domain walls are not expected to produce significant amounts of scalar radiation, except in the final stages of collapse  \cite{Vachaspati:1984yi}.  

These simple facts led us to consider a simplified domain wall model where the universe is permeated by a network of either spherical or parallel cylindrical domain walls with a given distribution of radii. Every cosmologically relevant domain wall is assumed to have started at rest at some early conformal time $\eta_i$ with an initial comoving radius $q_i$ much larger than the comoving horizon at that time ($\eta_i$). This condition  guarantees that in our model the evolution of every domain wall is that given by Eq. (\ref{dwev}) until the final collapse --- domain walls never intersect and the gravitational interaction between the walls is assumed to be negligible, thus  ensuring that they maintain the initial symmetry throughout the whole evolution. We assume that the walls decay when they reach $q=0$, which implies that at any given conformal time $\eta \gg \eta_i$ all domain walls with an initial comoving radius smaller than a given threshold $q_{i*}[\eta] \gg \eta_i$ would no longer be part of the network. We shall further  assume that the probability distribution for the initial radius of the domain walls is given by ${\mathcal P}(q_i) \propto q_i^{-2-s}$, so that the initial energy density of domain walls with $q_i$ larger than  $q_{i*} \gg \eta_i$ satisfies
\be
\rho_{wi} [q_i>q_{i*}] \propto \int_{q_{i*}}^\infty P(q_i) q_i^s dq_i \propto q_{i*}^{-1}\,.
\label{rhodwi}
\ee
For the sake of definiteness, we shall write the probability density function of the initial comoving radii as
\be
{\mathcal P} \equiv {\mathcal P} [q_i] =(1+s) \eta_i^{1+s} q_i^{-2-s} \Theta \left[q_i-\eta_i \right] \,,
\ee
where $\Theta$ is the Heaviside step function, thus ensuring that it is properly normalized ($\int_0^\infty {\mathcal P}(q_i) dq_i =1$) --- any other properly normalized probability density function consistent with Eq. (\ref{rhodwi}) would produce the same results.

\begin{figure}[!htb]
	\centering
	\includegraphics[width=8.0cm]{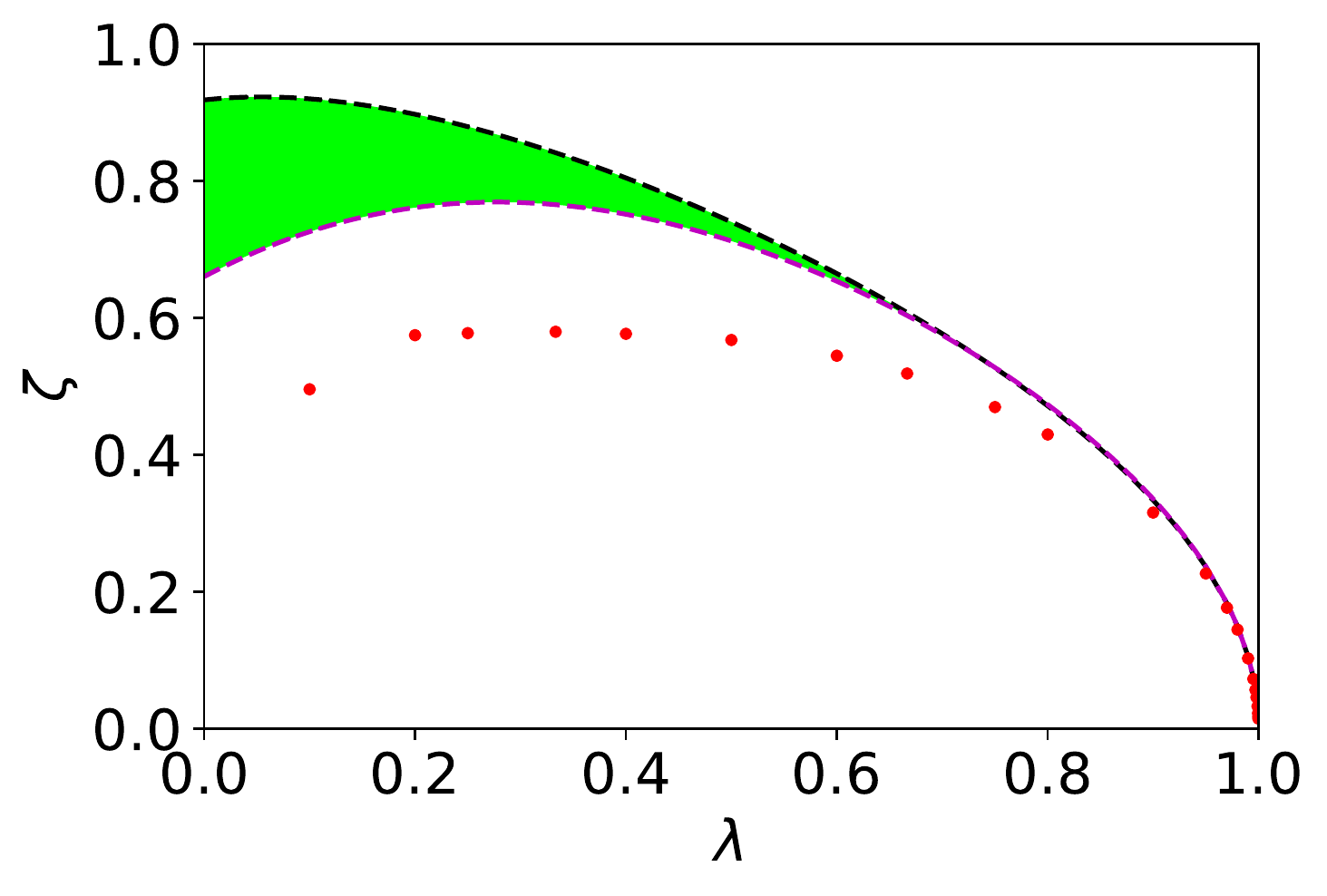}
	\caption{The value of $\zeta \equiv L/(a\eta)$ predicted by our parameter-free VOS model as a function of $\lambda$ considering cylindrical and spherical domain walls (upper black and lower magenta dashed lines, respectively) --- the green region between the two lines provides an estimate of the model uncertainty. The red dots represent results obtained using field theory numerical simulations of domain wall network evolution \cite{Martins:2016ois,Rybak2018}. Notice the exceptional agreement between the model predictions and the simulation results for $\lambda \ge 0.9$ and the significant discrepancies at smaller values of $\lambda$.}
	\label{fig3}
\end{figure}

As shown in Sec. \ref{sec2}, the wall energy (spherical domain wall) or energy per unit length (cylindrical domain wall) at the conformal time $\eta$ is given by
\bq
E&=&2\pi s \sigma_{w0} R^s = 2 \pi s \sigma_{w0} a^s q^s[\eta,q_i] \gamma [\eta,q_i] \nonumber \\
&=& 2 \pi s \sigma_{w0}  a^s \eta^{s} \tau^{-s} {\widetilde q}^s[\tau] \gamma[\tau] \,.
\eq
Hence, in our model the average macroscopic energy density of the domain wall network at the conformal time $\eta$ is equal to
\bq
\rho_{w}&=& \int _{q_{i*}}^\infty n E {\mathcal P} [q_i] dq_i  = a^{-1-s} \int _{q_{i*}}^\infty n_i E {\mathcal P} [q_i] dq_i \nonumber \\
&=& \sigma_{w0} \beta (a  \eta)^{-1} \int_0^{\tau_*} {\widetilde q}^s[\tau] \gamma[\tau] d \tau\,,
\eq
where $\beta=2 \pi s (1+s)  n_i \eta_i^{1+s}$, $n_i$ is the initial domain wall number density defined as the number of walls per unit volume (spherical case) or per unit area (cylindrical case), and $n= n_i a^{-1-s}$ would be the domain wall number density at the time $\eta$ in the absence of decay. {\tiny }The characteristic length of the network defined by $L \equiv \sigma_{w0}/\rho_w$ then satisfies 
\be
\zeta \equiv \frac{L}{a\eta}=\left(\beta \int_0^{\tau_*} {\widetilde q}^s[\tau] \gamma[\tau] d \tau\right)^{-1}\,.
\ee

\begin{figure}[!htb]
	\centering
	\includegraphics[width=8.0cm]{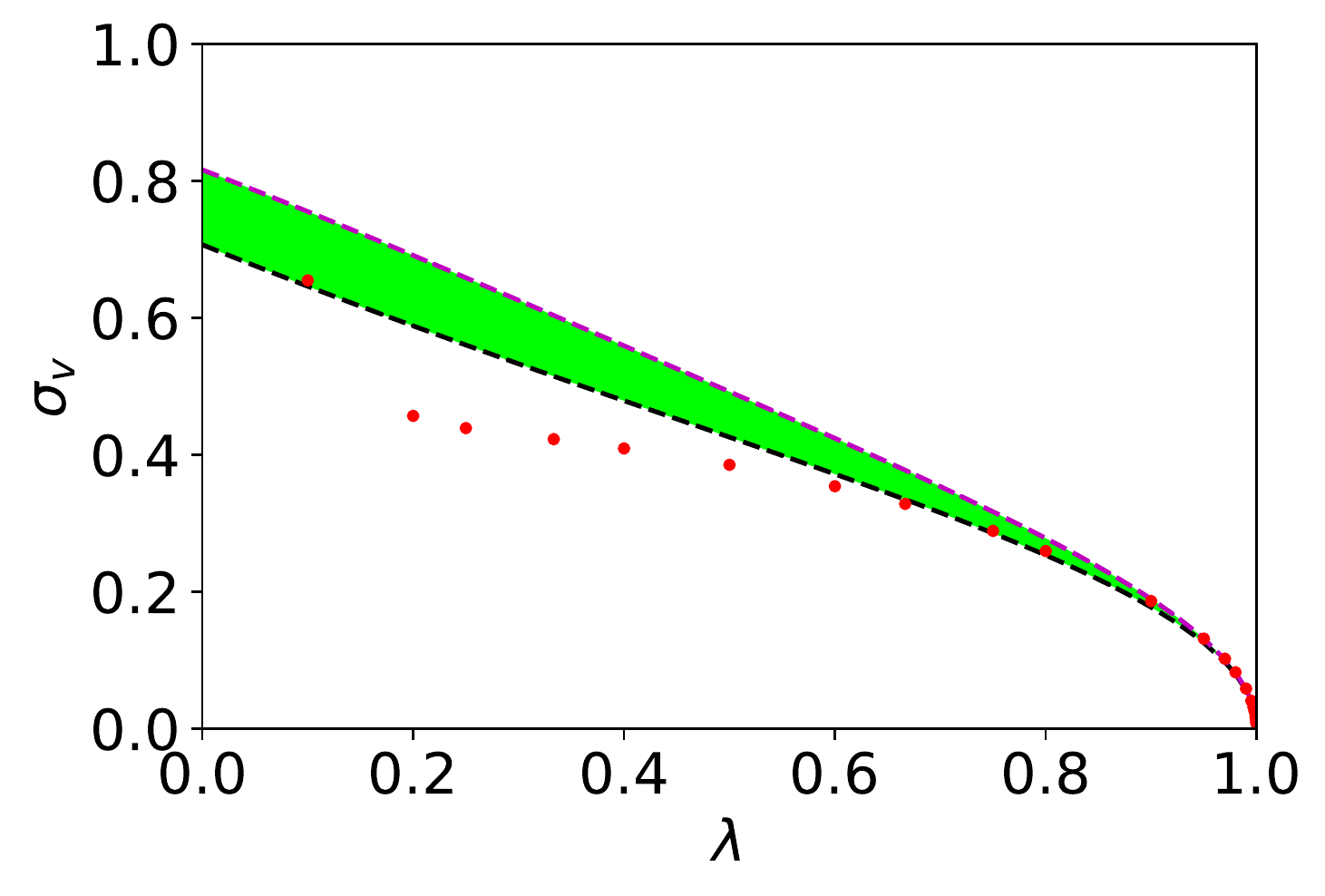}
	\caption{The value of root-mean square velocity $\sigma_v$ predicted by our parameter-free VOS model as a function of $\lambda$ considering cylindrical and spherical domain wall networks (lower black and upper magenta dashed lines, respectively) --- the green region between the two lines provides an estimate of the model uncertainty. The red dots represent results obtained using field theory numerical simulations of domain wall network evolution \cite{Martins:2016ois,Rybak2018}. Notice the exceptional agreement between the model predictions and the simulation results for $\lambda \ge 0.6$ and the significant discrepancies at smaller values of $\lambda$.}
	\label{fig4}
\end{figure}

Figure \ref{fig3} displays the value of $\zeta \equiv L/(a\eta)$ predicted by our parameter-free VOS model as a function of $\lambda$ considering cylindrical and spherical domain walls (upper black and lower magenta dashed lines, respectively). The green region between the two lines represents an estimate of the uncertainty associated to the geometry of the domain walls. Figure \ref{fig3} shows that this uncertainty is extremely small for values of $\lambda$ close to unity, growing significantly for smaller values of $\lambda$. Note that our model predicts the value of $\zeta$ up to a $\beta$-dependent normalization which will be discussed in the following section.

The mean-squared velocity of the domain walls may be computed as 
\bq
\sigma_v^2 &=& \frac{\int v^2 \rho dV}{\int \rho dV}=  \frac{\int v^2  \gamma dS}{\int  \gamma dS}= \frac{\int _{q_{i*}}^\infty v^2 E {\mathcal P} [q_i] dq_i }{\int _{q_{i*}}^\infty E {\mathcal P} [q_i] dq_i } \nonumber\\
&=&\frac{\int_0^{\tau_*} v^2[\tau] {\widetilde q}^s[\tau] \gamma[\tau] d \tau}{\int_0^{\tau_*} {\widetilde q}^s[\tau] \gamma[\tau] d \tau}\,,
\eq
where $\rho$ is the (microscopic) domain wall energy density at each point, $V$ is the physical volume, and $S$ is the domain wall area. The second equality is obtained by writing $dV=dSdl$ and performing the integration $\int \rho dl = \sigma_w=\sigma_{w0} \gamma$ in the direction perpendicular to the domain wall (note that $\rho=\rho_0 \gamma^2$ and $\delta=\delta_0/\gamma$, where the subscript `$0$' represents the proper rest value and $\delta$ denotes the domain wall thickness, and that $v$ does not vary along the direction perpendicular to the wall). Fig. \ref{fig4} shows the value of the root-mean square velocity $\sigma_v$ predicted by our parameter-free VOS model as a function of $\lambda$ considering cylindrical and spherical domain wall networks (lower black and upper magenta dashed lines, respectively). Again the green region between the two lines represents our estimate of the uncertainty associated to the geometrical properties of the domain wall network, by considering the two extreme domain wall geometrical configurations.

\section{Comparison with numerical simulations}
\label{sec4}

In this section we compare the predictions of our parameter-free VOS model with recent results of field theory numerical simulations of domain wall evolution  \cite{Martins:2016ois,Rybak2018}. The corresponding values of $\zeta$ as a function of $\lambda$ are represented by the red dots in Fig. \ref{fig3} --- the errors quoted by the authors being comparable to or smaller (in some cases  much smaller) than the size of the red dots. The value of $\beta$ of our parameter-free VOS model walls was fixed, in the cylindrical and spherical domain wall cases, by requiring the model to reproduce the simulation results for $\lambda=0.9998$, with $\beta_{\rm cylindrical} \sim 0.69$ and $\beta_{\rm spherical} \sim 1.15$ --- note that $\beta$ is independent of the cosmological model and, therefore, it is not a dynamical parameter of our model. Figure \ref{fig3} shows that the VOS model and the numerical results agree well in the non-relativistic limit, the discrepancies being smaller than $5 \%$ for $\lambda \ge 0.9$, but that for smaller  values of $\lambda$ the discrepancies are significant. 

In \cite{Martins:2016ois,Rybak2018} the authors quote the root-mean-square $\gamma v$ obtained from field theory numerical simulations of domain wall evolution (rather than $\sigma_v$) for various values of $\lambda$ in the interval $[0.1,0.9998]$. Furthermore, they compute it by calculating the average over the lattice volume of $(\gamma v)^2$, hence providing an estimate of
\bq
{\widehat \sigma}_{\gamma v}^2&=& \frac{\int (\gamma v)^2 dV}{\int dV}=  \frac{\int v^2  \gamma dS}{\int  \gamma^{-1} dS}=  \frac{\int _{q_{i*}}^\infty v^2 E {\mathcal P} [q_i] dq_i }{\int _{q_{i*}}^\infty \gamma^{-2} E {\mathcal P} [q_i] dq_i }\nonumber\\
&=&  \frac{\int _{q_{i*}}^\infty v^2 E {\mathcal P} [q_i] dq_i }{\int _{q_{i*}}^\infty E {\mathcal P} [q_i] dq_i }  \frac{\int _{q_{i*}}^\infty  E {\mathcal P} [q_i] dq_i }{\int _{q_{i*}}^\infty \gamma^{-2} E {\mathcal P} [q_i] dq_i }= \nonumber\\
&=& \sigma_v^2 (1+{\widehat \sigma}_{\gamma v}^2)
\,.
\eq
Here, we use the relation
\be
\sigma_v= \frac{{\widehat \sigma}_{\gamma v}}{\sqrt{1+{\widehat \sigma}_{\gamma v}^2}}\,,
\ee
in order to calculate the values of $\sigma_v$ from the quoted values of ${\widehat \sigma}_v$ estimated in \cite{Martins:2016ois,Rybak2018} using field theory numerical simulations of domain wall network evolution. These are represented by the red dots, whose radius is again typically of the same order or smaller (in some cases much smaller) than the quoted error bars. Figure \ref{fig4} shows that, in the non-relativistic regime, the values of $\sigma_v$ predicted by our parameter-free VOS model agree well with those obtained using field theory numerical simulations (the discrepancies being smaller than $5 \%$ for $\lambda \ge 0.6$) but that for larger values of $\lambda$ the differences are significant. 

\begin{figure}[!htb]
	\centering
	\includegraphics[width=8.0cm]{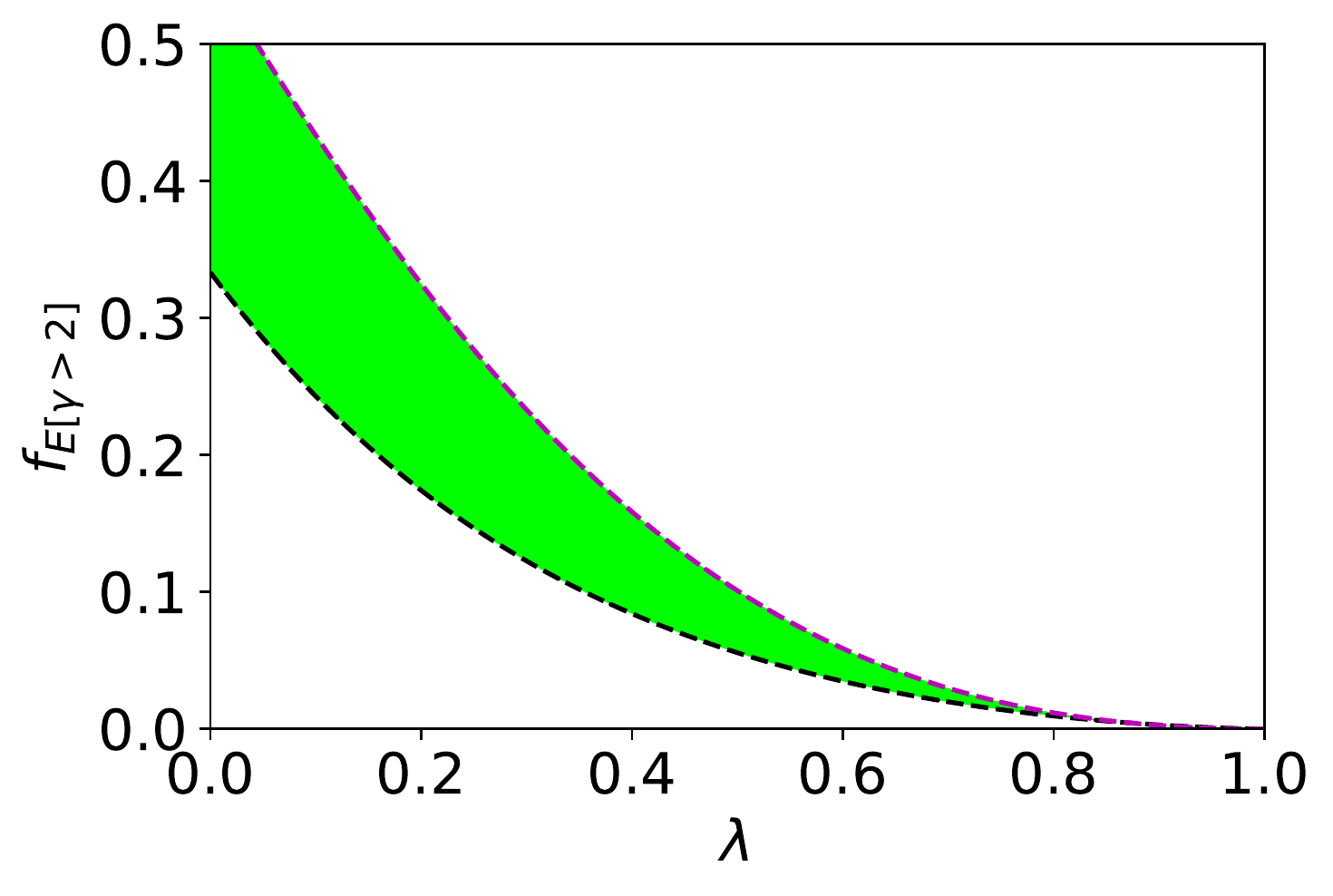}
	\caption{Fraction $f_{E[\gamma > 2]}$ of the energy of the domain wall network which is in the form of domain walls with $\gamma$ larger than $2$ as a function of $\lambda$. This fraction was computed using our parameter-free VOS model considering cylindrical and spherical domain walls (lower black and upper magenta dashed lines, respectively). Notice that, in both cases, $f_{E[\gamma>2]}$ deviates significantly from unity for values of $\lambda$ much smaller than unity.}
	\label{fig5}
\end{figure}

There are several possible causes for these differences. One of them is associated to the fact that, in order to produce accurate results, field theory numerical simulations of domain wall evolution need to be able resolve the domain walls. This is usually done by requiring the proper domain wall thickness of the domain walls to be significantly larger than the grid spacing used in the numerical simulations. Although, this is fine for non-relativistic domain walls, it might not be enough in the relativistic limit --- especially if $\gamma$ is significantly larger than unity --- due to the Lorentz contraction of the domain wall thickness. Figure \ref{fig5} shows the fraction $f_{E[\gamma > 2]}$ of the energy of the domain wall network in the form of domain walls with $\gamma$ greater than $2$ predicted by our new VOS model, showing that it becomes increasingly significant as $\lambda$ approaches zero. This effect, which has not yet been properly account for in field theory numerical simulations of domain wall network evolution, may help explaining the discrepancies between the predictions of our parameter-free VOS model and the results of field theory numerical simulations for values of $\lambda$ significantly smaller than unity.

In \cite{Martins:2016ois} it is argued, based on parametric fitting of a VOS model with six phenomenological parameters, that the energy losses due to the creation of sphere-like objects are typically subdominant in comparison to the contribution of scalar radiation. However, no significant scalar radiation is expected as long if the curvature radii associated with perturbations to the wall surface are always much larger than the domain wall thickness \cite{Vachaspati:1984yi}. Hence, domain walls are expected to emit significant amounts of scalar radiation only in the final stages of collapse. It is, however, possible that the artificially large domain wall thickness used in field theory numerical simulations in order to resolve the walls may lead to artificial  contributions to the scalar radiation component. Furthermore, as mentioned in the previous sections, the creation of sphere-like objects through wall intersection is not required in order for the network to lose energy as a result of domain wall collapse since a standard domain wall network is, from the very beginning, a collection of closed domain walls. The fast collapse of each of these walls begins roughly when its comoving characteristic size equals ${\mathcal H}^{-1}$, thus leading to a hierarchical collapse starting with the smallest domain walls.

Another possible contribution to the differences between the predictions of our VOS model and those based on field theory numerical simulations has to do with the fact that, as $\lambda$ approaches zero, the fractional contribution of the scalar radiation to the total energy associated to the scalar field becomes increasingly larger in field theory simulations. This might lead to spurious contributions to the estimated values of $L$ and $\sigma_v$, especially in the relativistic limit.

\section{Conclusions}
\label{sec5}

In this paper we have developed a new parameter-free VOS model for the evolution of the characteristic length and root-mean-square velocity of domain wall networks in arbitrary FLRW backgrounds. This model incorporates the main physical ingredients relevant for the determination of the evolution of the macroscopic properties of domain wall networks. We have shown that the predictions of our model are consistent with the results of field theory numerical simulations of domain wall evolution in the non-relativistic limit. However, away from this limit we have found significant discrepancies between the predictions of our new VOS model and the results obtained from numerical simulations. We have identified possible causes of these discrepancies, mainly associated to the determination of the characteristic length and the root-mean-square velocity of domain wall networks from numerical simulations, which will need to be tackled in future work. Although the results of the present paper are not directly applicable to other defect networks --- for example, cosmic strings loops may intersect frequently and oscillate several times before decaying --- this research might inspire other works in the same spirit, that could  contribute to the further development of accurate semi-analytical models for cosmic defect network evolution with a minimum number of phenomenological parameters.

\begin{acknowledgments}
	
P. P. A. thanks the support from FCT---Funda{\c c}\~ao para a Ci\^encia e a Tecnologia through the Sabbatical Grant No. SFRH/BSAB/150322/2019. P. P. A. also acknowledges the support by FEDER—Fundo Europeu de Desenvolvimento Regional funds through the COMPETE 2020---Operational Programme for Competitiveness and Internationalisation (POCI) and by Portuguese funds through FCT in the framework of Project No. POCI-01-0145-FEDER-031938. Funding of this work has also been provided by the FCT Grant No. UID/FIS/04434/2019.
\end{acknowledgments}
 
\bibliography{walls}
 	
 \end{document}